\documentclass[10pt]{iopart}

\expandafter\let\csname equation*\endcsname\relax
\expandafter\let\csname endequation*\endcsname\relax
\usepackage{amsmath}
\usepackage{amssymb}
\usepackage{bbm}
\usepackage{color}
\usepackage{graphicx}
\usepackage{hyperref}
\usepackage{iopams}
\usepackage[square,numbers,sort&compress]{natbib}
\usepackage{physics}
\usepackage{tikz}
\usepackage{soul}

\usetikzlibrary{graphs,graphs.standard,quotes,angles,backgrounds,shapes,snakes}
\definecolor{vertexblue}{HTML}{a3b4cc}
\tikzstyle{vert} = [circle,minimum size=6mm,draw=black,fill=vertexblue, line width=.5mm]
\tikzstyle{edg} = [red, line width=1mm]

\newlength{\gwidth} 
\hypersetup{
  colorlinks=true,linkcolor=blue,citecolor=blue,
  filecolor=blue,urlcolor=blue,breaklinks=true
}

\newcommand{\orcid}[1]{\href{https://orcid.org/#1}
{\includegraphics[width=7pt]{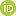}}}

\def\be{\begin{equation}}
\def\ee{\end{equation}}

\def\bc{\begin{center}}
\def\ec{\end{center}}
\def\bal{\begin{align}}
\def\eal{\end{align}}

\begin{document}

\title{Entangled states from simple quantum graphs}

\author{Alison A. Silva\orcid{0000-0003-3552-8780}}
\ead{alisonantunessilva@gmail.com}
\address{
  Programa de Pós-Graduação em Ciências/Física,
  Universidade Estadual de Ponta Grossa,
  84030-900 Ponta Grossa, Paraná, Brazil
}

\author{D. Bazeia\orcid{0000-0003-1335-3705}}
\ead{bazeia@fisica.ufpb.br}
\address{
  Departamento de Física,
  Universidade Federal da Paraíba,
  58051-900 João Pessoa, Paraíba, Brazil
}

\author{Fabiano M. Andrade\footnote{Corresponding author}\orcid{0000-0001-5383-6168}}
\ead{fmandrade@uepg.br}
\address{
  Programa de Pós-Graduação em Ciências/Física,
  Universidade Estadual de Ponta Grossa,
  84030-900 Ponta Grossa, Paraná, Brazil
}
\address{
  Departamento de Matemática e Estatística,
  Universidade Estadual de Ponta Grossa,
  84030-900 Ponta Grossa, Paraná, Brazil
}
\address{
  Departamento de Física,
  Universidade Federal do Paraná,
  81531-980 Curitiba, Paran\'a, Brazil
}

\date{\today}

\begin{abstract}
Entanglement is a fundamental resource for many applications in quantum information processing. Here, we investigate how quantum transport in simple quantum graphs, modeled as controlled two-level quantum systems, can be utilized to generate entangled states through coherent control operations between two simple quantum graphs. A controlled operation is defined such that the scattering behavior of one quantum graph dynamically modifies the other. Our analysis reveals the precise conditions under which maximal entanglement or separability arises, including configurations that can be implemented via phase shifts in graph structures. Our findings demonstrate that the maximal entanglement in this system is closely related to recent results on randomized quantum graphs. These results provide new pathways for engineering entanglement using simple quantum graphs and suggest experimental feasibility using microwave networks.\\

\noindent doi: \href{https://doi.org/10.1088/1751-8121/ae1270}
{10.1088/1751-8121/ae1270}
\end{abstract}

\maketitle

\section{Introduction}
Quantum graphs (QGs) \cite{Book.Berkolaiko.2006,Book.Berkolaiko.2013}
have been employed to describe different phenomena in several areas of
science.
In particular, they may arise as simplified models in chemistry,
mathematics, physics, and have become a powerful tool for studying
different aspects of the natural world.
Specific subjects include the study of quantum chaos
\cite{PRL.79.4794.1997,AoP.274.76.1999}, Anderson localization
\cite{PRL.84.1427.2000}, and chaotic and diffusive scattering
\cite{PRL.85.968.2000,PRE.65.016205.2001}, to name just a few
possibilities of current interest.

In this Letter, we focus mainly on a novel investigation concerning the
construction of a direct connection between QGs and quantum systems
within the quantum information framework.
To be more specific, we shall deal with scattering processes in QGs
equipped with two scattering channels and their relation with a
two-level quantum system.
This is perhaps the most straightforward possibility, and extensions
involving three or more states will be considered elsewhere.
The purpose of this study is to relate the properties of the quantum
states generated by a given QG and their possible connections to the
two-level quantum states they generate.
The construction allows us to define controlled operations
between the QGs, which will be used to infer properties of the two-level
quantum system on the quantum information side.
We introduce a model connecting open QGs to entangled two-level quantum
systems, advancing the previous work
\cite{JMS.153.197.2008,JPA.54.445202.2021}
by defining a coherent control operation between two QGs.
Unlike previous treatments, which employ static S-matrix formalism for
quantum computation, our method generates entanglement conditioned on
graph topology and tunable physical parameters such as edge lengths or
phase shifts.

This framework is not purely abstract, as recent developments in
microwave network analogs of QGs
\cite{PRE.108.034219.2023,PRE.107.024203.2023,PRE.104.045211.2021,
PRE.69.056205.2004}
suggest that controlled phase shifts and edge manipulations are
experimentally accessible.
Our simple models involving only two channels and single-edge graphs are
particularly amenable to implementation using tunable waveguides and
phase shifters.

\section{Quantum graphs}
A graph $G=(V,E)$ is defined as a pair consisting
of a finite set of vertices $V=\{v_1,\ldots,v_n\}$ and a set of edges
$E=\{e_1,\ldots,e_l\}$ \cite{Book.2010.Diestel}.
A QG is a triple $\{\Gamma_{G},H,\text{BC}\}$ consisting of a metric
graph $\Gamma_{G}$, a graph $G$ with positive lengths
$\ell_{e_{i,j}}\in (0,\infty)$ on each edge $\{i,j\}$, a differential
operator $H$, and a set of boundary conditions (BCs) at the vertices,
which define the individual scattering amplitudes at the vertices
\cite{Book.Berkolaiko.2013}.
We consider the stationary free Schr\"odinger operator
$H=-(\hbar^2/2m)d^2/dx^2$ on each edge.
Then, to create an open QG with two scattering channels, $\Gamma_{G}^2$,
we add two leads (semi-infinite edges) to two different vertices.
The open QG $\Gamma_G^2$ is characterized by the global energy-dependent
scattering matrix $\sigma_{\Gamma_G^2}^{(f,i)}(k)$, where, as usual,
$k=\sqrt{2mE/\hbar^2}$ is the wave number, with $i$ and $f$ the entrance
and exit scattering channels, respectively.
Thus, the global scattering amplitudes are intrinsically determined by
the graph topology, its metric structure, the BCs at the vertices, and
the energy of the incoming particle.
This rich interplay of parameters provides a high degree of control over
the scattering behavior, establishing quantum graphs as versatile and
tunable platforms for investigating scattering phenomena.

In general, for an open QG with $c$ scattering channels, we can obtain
the global scattering amplitudes $\sigma_{\Gamma_G^c}^{(f,i)}(k)$ by
employing the Green's function approach, which was developed in Refs. \cite{PR.647.1.2016,PRA.98.062107.2018,JPA.56.475202.2023}.
This approach is general and has been successfully used in several
transport studies in QGs \cite{PRA.103.062208.2021,APQ.1.046126.2024}.
For a QG $\Gamma_G^{2}$ with two scattering channels (represented by
$\Gamma_G$ to simplify notation), labeled $l_0$ and $l_1$, the
scattering $S$-matrix is defined in terms of the global scattering
amplitudes
\begin{equation}
  S_{\Gamma_{G}}=
  \begin{pmatrix}
    r_G & t_G\\
    t_G & r_G
  \end{pmatrix}.
\end{equation}
where $r_{G}$ and $t_G$ are the global scattering amplitudes of the QG
$\Gamma_{G}$.

\begin{figure}[t]
  \centering
  \includegraphics[width=.6\linewidth]{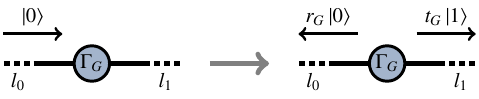}
  \caption{
    The scattering in QGs as an information model.
    Initially, there is an input state in a given QG $\Gamma_G$, where a
    particle enters by one of its leads, called $l_0$, and thus the
    particle is found to be in the state $\ket{0}$.
    After the scattering in $\Gamma_G$, there is an output state where
    the particle has two scattering amplitudes: $r_G$ to reflect to the
    same lead $l_0$ and thus it is in the state $\ket{0}$; and $t_G$ to
    transmit through the QG and be found in the other lead $l_1$, being
    in the state $\ket{1}$.
  }
  \label{fig:fig1}
\end{figure}

\section{Quantum scattering states}
We model the two QG scattering channels as quantum states $\ket{0}$ and
$\ket{1}$.
Defining an initial state as a signal entering from the lead labeled
$l_0$, corresponding to the input state $\ket{\psi_i} = \ket{0}$.
After the scattering process in $\Gamma_G$, with the scattering matrix
$S_{\Gamma_G}$, the output state is
$\ket{\Gamma_G} = S_{\Gamma_G} \ket{0} = r_G \ket{0} + t_G \ket{1}$.
This is illustrated in Fig. \ref{fig:fig1}.

\begin{figure}[t]
  \centering
  \includegraphics[width=.5\linewidth]{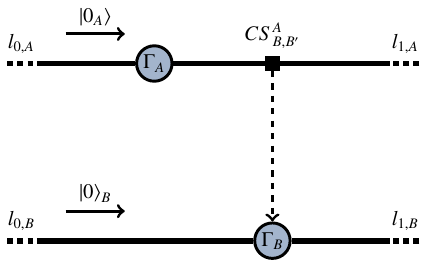}
  \caption{
    The coherent control operation between two QGs $\Gamma_A$ and
    $\Gamma_B$ by the controlled scattering operator, which may change
    $\Gamma_B$ to $\Gamma_{B'}$.
  }
  \label{fig:fig2}
\end{figure}

Let us introduce $\Gamma_{A}$ as Alice's QG and $\Gamma_{B}$ as
Bob's QG.
Thus, the state associated with the scattering in each QG is defined by
$\ket{\Gamma_i} = S_i \ket{0_i}$ with $i = A,B$.
We now define a \emph{controlled scattering operator}
\begin{equation}
  CS_{B,B'}^{A} = \ketbra{0_{A}}\otimes S_{\Gamma_B}
               + \ketbra{1_{A}}\otimes S_{\Gamma_{B'}},
\end{equation}
analogous to the standard controlled unitary operation in quantum
computing.
This operator captures how Alice’s scattering outcome conditions Bob’s
QG, modifying $\Gamma_B$ to $\Gamma_{B'}$ by a change of parameters in
the same QG or even changing it to another QG.
Alice's QG acts as the control, while Bob's QG is the one being
controlled.
This form of control mimics a quantum logic gate, such as controlled-$Z$
or controlled-phase, with the advantage of being directly realized via
modifiable QG parameters.
These modifications may include changes in the edge lengths, potentials
applied on its edges, in the boundary conditions at the vertices, etc.
This is illustrated in Fig. \ref{fig:fig2}.
Thus, the global state can be written as
\begin{equation}
  \ket{\Psi_{AB}}
  = CS_{B,B'}^{A} (S_{\Gamma_A}
    \otimes \mathbbm{1}) \ket{0_A} \otimes \ket{0_B}
  = r_{A} \ket{0_A} \otimes \ket{\Gamma_B}
    + t_{A} \ket{1_A}\otimes\ket{\Gamma_{B'}},
\end{equation}
where $\mathbbm{1}$ is the $2\times2$ identity matrix.
When $\ket{\Gamma_{B'}} = \ket{\Gamma_{B}}$, the state is separable as
$\ket{\Psi_{AB}} = \ket{\Gamma_A}\otimes\ket{\Gamma_B}$.
Using the shorthand notation $\ket{i_A} \otimes \ket{j_B} = \ket{i,j}$,
we can write the global scattering state as
\begin{equation}
  \ket{\Psi_{AB}} =
  r_{A} r_{B}  \ket{0,0} + r_{A} t_{B}  \ket{0,1} +
  t_{A} r_{B'} \ket{1,0} + t_{A} t_{B'} \ket{1,1}.
  \label{eq:ketPhi}
\end{equation}
The related global density matrix $\rho_{AB}=\ketbra{\Psi_{AB}}$ has the
form
\begin{equation}
  \rho_{AB} =
    \begin{pmatrix}
      |r_{A}|^{2}|r_{B}|^{2} & |r_{A}|^{2}r_{B}t_{B}^{*}
      & r_{A}r_{B}t_{A}^{*}r_{B'}^{*} & r_{A}r_{B}t_{A}^{*}t_{B'}^{*}\\
      |r_{A}|^{2}t_{B}r_{B}^{*} & |r_{A}|^{2}|t_{B}|^{2}
      & r_{A}t_{B}t_{A}^{*}r_{B'}^{*} & r_{A}t_{B}t_{A}^{*}t_{B'}^{*}\\
      t_{A}r_{B'}r_{A}^{*}r_{B}^{*} & t_{A}r_{B'}r_{A}^{*}t_{B}^{*}
      & |t_{A}|^{2}|r_{B'}|^{2} & |t_{A}|^{2}r_{B'}t_{B'}^{*}\\
      t_{A}t_{B'}r_{A}^{*}r_{B}^{*} & t_{A}t_{B'}r_{A}^{*}t_{B}^{*}
      & |t_{A}|^{2}t_{B'}r_{B'}^{*} & |t_{A}|^{2}|t_{B'}|^{2}
\end{pmatrix}.
\end{equation}

Alice's state is obtained by taking the partial trace over Bob's
subsystem, resulting in
\begin{equation}
  \rho_{A}=
  \tr_{B}(\rho_{AB})=
  \begin{pmatrix}
    \left|r_{A}\right|^{2}
    & r_{A}t_{A}^{*}\left(r_{B}r_{B'}^{*}+t_{B}t_{B'}^{*}\right)\\
    t_{A}r_{A}^{*}\left(r_{B'}r_{B}^{*}+t_{B'}t_{B}^{*}\right)
    & \left|t_{A}\right|^{2}
  \end{pmatrix}.
  \label{eq:partialtraceAlice}
\end{equation}
The diagonal elements of the reduced density matrix correspond to the
expected scattering probabilities.
In this sense, measurements in Alice's QG scattering channels return
the reflection and transmission probabilities of this QG, i.e.
$\left|r_{A}\right|^{2}$ and $\left|t_{A}\right|^{2}$.
On the other hand, if one takes the partial trace over Alice's
subsystem, we obtain Bob's reduced density matrix
\begin{equation}
  \rho_{B}=
  \tr_{A}(\rho_{AB})=
  \begin{pmatrix}
    |r_{A}|^{2}|r_{B}|^{2}+|t_{A}|^{2}|r_{B'}|^{2}
    & |r_{A}|^{2}r_{B}t_{B}^{*}+|t_{A}|^{2}r_{B'}t_{B'}^{*}\\
    |r_{A}|^{2}t_{B}r_{B}^{*}+|t_{A}|^{2}t_{B'}r_{B'}^{*}
    & |r_{A}|^{2}|t_{B}|^{2}+|t_{A}|^{2}|t_{B'}|^{2}
  \end{pmatrix}.
\end{equation}
We can use this result to obtain the expected transmission value in
Bob's QG as
$\bra{1}\rho_{B}\ket{1}=|r_{A}|^{2}|t_{B}|^{2}+|t_{A}|^{2}|t_{B'}|^{2}$.
By introducing a randomization parameter $p$, defined as
$p=|t_{A}|^{2}$, the expression can be rewritten as
\begin{equation}\label{rqg}
   \bra{1}\rho_{B}\ket{1}=\left(1-p\right)|t_{B}|^{2}+p|t_{B'}|^{2}.
\end{equation}
This shows that measurements in Bob's QG scattering channels can be
interpreted as a linear combination of the scattering probabilities
associated with the two distinct configurations of the system.
A related approach was recently proposed in
Ref. \cite{APQ.1.046126.2024}, where the parameter $p$ was introduced to
model the probability of the existence of an edge within a single QG.
In contrast, our interpretation assigns $p$ a dynamical role,
characterizing the coherent control between Alice’s and Bob’s QGs.
This distinction suggests that our result may offer broader practical
relevance as it extends the notion of randomization from structural
uncertainty to control between the two graphs.

\begin{figure}[b]
    \centering
    \includegraphics[width=.5\linewidth]{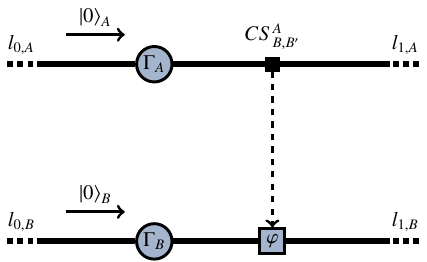}
    \caption{
    The controlled scattering system of two QGs $\Gamma_A$ and
    $\Gamma_B$, where the controlled operation is a phase $\varphi$
    applied to the scattering channel $l_{1,B}$ by a controlled
    scattering operator $CS_{B,B'}^{A}$ in the scattering channel
    $l_{1,A}$.}
    \label{fig:fig3}
\end{figure}

\section{Entanglement entropy}
The definition of the von Neumann entropy related to the density matrix
$\rho$ is given by $H(\rho) = -\tr\left(\rho \log_2 \rho\right)$.
When applied to the reduced density matrix $\rho_A$ of a system formed
by two subsystems $A$ and $B$, one can follow Refs.  \cite{PRL.78.2275.1997,PRA.57.1619.1998,RMP.81.865.2009}
to introduce the entanglement entropy (EE) in the form
$H(\rho_{A}) = -\sum_j \lambda_j \log_2 \lambda_j$,
where $\lambda_j$ is the eigenvalue of $\rho_A$.
Thus, taking this entropy, with $A$ being a two-level system, with
eigenvalues \{$\lambda_+$,$\lambda_-$\}, we have
$H(\rho_{A}) = - \lambda_+ \log_2 \lambda_+ - \lambda_- \log_2 \lambda_-$.
This can be applied to the reduced density matrix from
Eq. \eqref{eq:partialtraceAlice} which has eigenvalues
\begin{equation}
  \lambda_{\pm}=
  \frac{1}{2}
  \left[
    1\pm
    \sqrt{
      1-
      4\left|r_{A}\right|^{2}\left|t_{A}\right|^{2}
      \left(1-\left|r_{B}r_{B'}^{*}+t_{B}t_{B'}^{*}\right|^{2}\right)
    }
  \right].
\end{equation}
These eigenvalues, together with the entanglement entropy, help us
verify the relationship between the scattering amplitudes and the
presence of entanglement in this system.

The EE is maximized when the eigenvalues are equal.
This imposes
\begin{equation}
  \sqrt{
    1-4\left|r_{A}\right|^{2}\left|t_{A}\right|^{2}
    \left(1-\left|r_{B}r_{B'}^{*}+t_{B}t_{B'}^{*}\right|^{2}\right)}=0
\end{equation}
which leads to
\begin{equation}
  \left|r_{A}\right|^{2}\left|t_{A}\right|^{2}
  \left(1-\left|r_{B}r_{B'}^{*}+t_{B}t_{B'}^{*}\right|^{2}\right)
  =\frac{1}{4}.
    \label{eq:maxentcond}
\end{equation}
The only solution is obtained for
$\left|r_{A}\right|^{2} = \left|t_{A}\right|^{2}=1/2$, and
 \begin{equation}\label{21}
\left|r_{B}r_{B'}^{*}+t_{B}t_{B'}^{*}\right|^{2}=0.
 \end{equation}
It shows that Alice's QG must have maximum scattering entropy
\cite{PRA.103.062208.2021}, while the restriction in Bob's QG given by
Eq. \eqref{21} indicates that to achieve maximum EE, the change in the
scattering amplitudes on $\Gamma_B$ should be such that
$\left|t_{B'}\right|^{2} = \left|r_{B}\right|^{2}$ and consequently
$\left|r_{B'}\right|^{2} = \left|t_{B}\right|^{2}$.
That is a swapping between its scattering probabilities.
Moreover, when
\begin{equation}    \left|r_{A}\right|^{2}\left|t_{A}\right|^{2}
  \left(1-\left|r_{B}r_{B'}^{*}+t_{B}t_{B'}^{*}\right|^{2}\right)=0,
  \label{eq:sepcond}
\end{equation}
the EE is zero, and the system is separable.
This is obtained if the transmission or the reflection probability is
zero in Alice's QG, or when
$\left|r_{B}r_{B'}^{*}+t_{B}t_{B'}^{*}\right|^{2} = 1$, which can be
obtained when there is no change in Bob's QG scattering amplitudes,
or when a global phase is applied, or even when a phase $2 \pi n$ is
added to one of these scattering amplitudes.
Thus, the EE of the system vanishes when the scattering entropy in
Alice's QG is zero, or if the modification in Bob's QG is equivalent to
a global phase.

\begin{figure}[t]
    \centering
    \includegraphics[height=.6\linewidth]{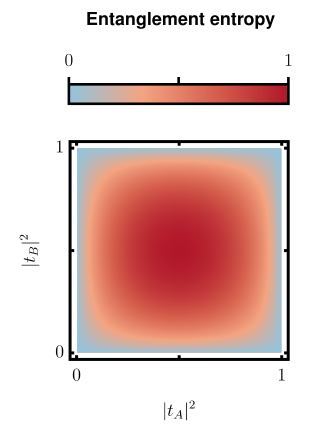}
    \caption{
    The EE as a function of the transmission probability $|t_A|^2$ and
    $|t_B|^2$ in each QG when a controlled phase $\pi$ is applied in
    Bob's QG, with $t_{B'}=-t_{B}$.}
    \label{fig:fig4}
\end{figure}

\section{Results}
Let us now include a controlled change in Bob's QG as a phase $\varphi$
applied to its transmission channel (see Fig. \ref{fig:fig3}).
In this way, Eq. \eqref{eq:ketPhi} becomes
\begin{equation}
  \ket{\Psi_{AB}} =
  r_{A}r_{B} \ket{0,0} + r_{A}t_{B} \ket{0,1}
  + t_{A}r_{B} \ket{1,0} + e^{i \varphi} t_{A}t_{B} \ket{1,1}.
\end{equation}
For the maximal entanglement obtained in Eq. \eqref{eq:maxentcond}, we
find that this new system may present a maximal entanglement when
\begin{equation}
  \left|r_{A}\right|^{2}\left|t_{A}\right|^{2}
  \left(1-\left|\left|r_{B}\right|^{2}+e^{-i \varphi}
      \left|t_{B}\right|^{2}\right|^{2}\right)
  =\frac{1}{4}.
\end{equation}
The solution for a real $\varphi$ is
$\left|r_{A}\right|^{2} = \left|t_{A}\right|^{2}=1/2$,
$\left|r_{B}\right|^{2} = \left|t_{B}\right|^{2}=1/2$ and
$\varphi= \left(2 n + 1\right) \pi$.
This is equivalent to applying a controlled-$Z$ gate together with phase
shifts.
Furthermore, from the separability condition obtained in
Eq. \eqref{eq:sepcond}, one notices that this new system will be
separable for
\begin{equation}
  \left|r_{A}\right|^{2}\left|t_{A}\right|^{2}
  \left(
    1-\left|\left|r_{B}\right|^{2}+
      e^{-i \varphi} \left|t_{B}\right|^{2}\right|^{2}
  \right)=0.
\end{equation}
This is obtained when $\left|r_{A}\right|^{2} = 0$, or
$\left|t_{A}\right|^{2} = 0$ or $\varphi=2 n \pi$.
The EE for this system for $\varphi=\pi$ is illustrated in
Fig. \ref{fig:fig4}.

\begin{figure}[t]
  \centering
  \includegraphics[width=.5\linewidth]{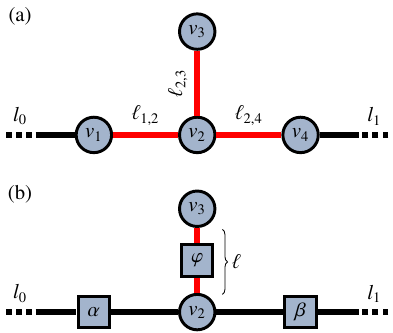}
  \caption{
    (a) A star QG with four vertices ($S_4$) and two scattering channels
    ($\Gamma_{S_4}$) which are connected to the vertices labeled as
    $v_1$ and $v_4$.
    Each edge which connects a pair of vertices $\{i,j\}$ in the QG has
    a given length $\ell_{i,j}$.
    (b) An equivalent QG where a phase shift operation $\alpha$ is used
    to mimic the phase obtained due to the length $\ell_{1,2}$, and the
    same is done to $\ell_{2,4}$, being replaced by $\beta$.
    The distance between the vertices $v_2$  and $v_3$ is $\ell$ and a
    possible phase shift operation along this edge is defined by
    $\varphi$.
  }
  \label{fig:fig5}
\end{figure}

We can also consider the case where a controlled phase acts inside the
QG.
To do that, let us now study the simple case of star quantum graphs.
First, we consider the scattering amplitudes in the QG presented in
Fig. \ref{fig:fig5} (a) with standard boundary conditions \cite{Incollection.2017.Berkolaiko,Book.Berkolaiko.2013}.
The corresponding scattering matrix is given by
\begin{equation}
  S_{\Gamma_{S_{4}}}=
  \frac{-1}{2i+\tan\left(k\ell_{2,3}\right)}
  \begin{pmatrix}
    \tan\left(k\ell_{2,3}\right)e^{2ik\ell_{1,2}}
    & -2ie^{ik\left(\ell_{1,2}+\ell_{2,4}\right)}\\
    -2ie^{ik\left(\ell_{1,2}+\ell_{2,4}\right)}
    & \tan\left(k\ell_{2,3}\right)e^{2ik\ell_{2,4}}
  \end{pmatrix}.
\end{equation}
Since we are considering the Neumann boundary condition, the scattering
amplitudes in the vertices $v_1$ and $v_4$ are $r=0$ and $t=1$.
Thus, since the lengths $\ell_{1,2}$ and $\ell_{2,4}$ contribute only
to the phase change in the scattering amplitudes, it is interesting to
simplify them to phase equivalent parameters $\alpha=k\ell_{1,2}$ and
$\beta=k\ell_{2,4}$.
In the same way, we can rewrite the length $\ell_{2,3}$ as simply $\ell$
and add a possible phase inclusion $\varphi$, obtaining
$k\ell_{2,3}=k\ell+\varphi$.
An illustration of these changes is shown in Fig. \ref{fig:fig5}(b).
Thus, using these modifications, the scattering matrix is given by
\begin{equation}
  S_{\Gamma_{S_{4}}}(\alpha,\beta,\varphi)=
  -\frac{1}{2i+\tan\left(k\ell+\varphi\right)}
  \begin{pmatrix}
    \tan\left(k\ell+\varphi\right)e^{2i\alpha}
    & -2ie^{i\left(\alpha+\beta\right)}\\
    -2ie^{i\left(\alpha+\beta\right)}
    & \tan\left(k\ell+\varphi\right)e^{2i\beta}
  \end{pmatrix}.
\end{equation}
Some examples of quantum logical gates
\cite{Book.2009.Barnett,Book.2010.Nielsen} that can be obtained through
the manipulation of the variables $\varphi$, $\alpha$ and $\beta$ as
variables are shown in Table \ref{tab}.
For simplicity, setting $\varphi=0$, $\alpha=\pi$, and $\beta= \pi$,
the scattering matrix takes the form
\begin{equation}
  S_{\Gamma_{S_{4}}}(0,\pi,\pi)=
  -\frac{1}{2i+\tan\left(k\ell\right)}
  \begin{pmatrix}
    \tan\left(k\ell\right) & -2i\\
    -2i & \tan\left(k\ell\right)
\end{pmatrix}.
\end{equation}
From the S-matrix above, we can extract the quantum amplitudes
\begin{equation}
  \mathcal{R}(k \ell) =
  -\frac{\tan\left(k\ell\right)}{2i+\tan\left(k\ell\right)},
  \text{ and }
  \mathcal{T}(k \ell) =
  \frac{2i}{2i+\tan\left(k\ell\right)}.
\end{equation}

\begin{table*}[b]
  \caption{
      Parameters in the quantum graph from Fig. \ref{fig:fig5} in order
      to simulate some quantum logical gates.}
  \begin{center}
    \label{tab}
    \begin{tabular}{cccc}
      \hline
      Quantum Logical Gate & $k\ell+\varphi$ & $\alpha$ & $\beta$\\
      \hline
      Identity & $\left(n_{\varphi}+\frac{1}{2}\right)\pi$ & $\left(n_{\alpha}+\frac{1}{2}\right)\pi$ & $\left(n_{\beta}+\frac{1}{2}\right)\pi$\\
      Global phase $\delta$ & $\left(n_{\varphi}+\frac{1}{2}\right)\pi$ & $\left(n_{\alpha}+\frac{1}{2}\right)\pi+\frac{\delta}{2}$ & $\left(n_{\beta}+\frac{1}{2}\right)\pi+\frac{\delta}{2}$\\
      Pauli X & $n_{\varphi}\pi$ & $2n_{\alpha}\pi-\beta$ & $2n_{\beta}\pi-\alpha$\\
      Pauli Z & $\left(n_{\varphi}+\frac{1}{2}\right)\pi$ & $\left(n_{\alpha}+\frac{1}{2}\right)\pi$ & $n_{\beta}\pi$\\
      Hadamard & $\pm\arctan\left(2\right)+n_{\varphi}\pi$ &
      $\left(n_{\alpha}-\frac{3}{8}\right)\pi$ & $\left(2n_{\beta}-n_{\alpha}\pm\frac{1}{8}\right)\pi$\\
      \hline
    \end{tabular}
  \end{center}
\end{table*}

\begin{figure}[t]
  \centering
  \includegraphics[width=.5\linewidth]{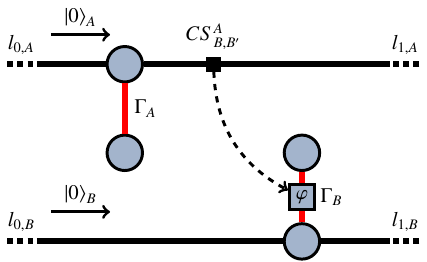}
  \caption{
    Illustration of the system composed of two QGs $\Gamma_A$ and
    $\Gamma_B$, where a controlled phase shift $\varphi$, activated by a
    controlled scattering operator $CS_{B,B'}^{A}$, is applied to the
    shown edge of $\Gamma_B$.}
  \label{fig:fig6}
\end{figure}

\begin{figure}[t]
  \centering
  \includegraphics[height=.65\linewidth]{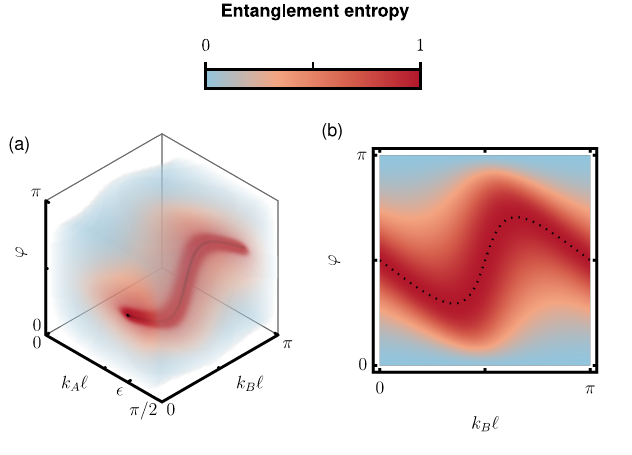}
  \caption{
    The EE in a system with two QGs, with wavenumbers
    $k_A$ and $k_B$, when a controlled phase $\varphi$ is applied in the
    edge of Bob's QG, as illustrated in Fig. \ref{fig:fig6}.
    (a) The case with the three variables $k_A \ell$, $k_B \ell$ and
    $\varphi$.
    (b) The case with $k_A\ell=\epsilon=\arctan{2}$, highlighting the
    maximal entropy by a black dotted line.
  }
  \label{fig:fig7}
\end{figure}

For the coherent control operation between two of these QGs, we suppose
that Alice's and Bob's QGs are initially two identical star QGs, with
four vertices, illustrated in Fig. \ref{fig:fig5}, each one with a wave
number $k_A$ and $k_B$.
Moreover, we introduce a controlled operation in Bob's QG as an
additional phase $\varphi$ operation along the edge $e_{2,3}$.
This is illustrated in Fig. \ref{fig:fig6}.
Thus, the scattering state in this system is given by
\begin{align}
  \label{eq:scatteringstateRsandTs}
  \ket{\Psi_{AB}} = {}
  &
    \mathcal{R}(k_A \ell)\mathcal{R}(k_B \ell)\ket{0,0}
    +\mathcal{R}(k_A \ell)\mathcal{T}(k_B \ell)\ket{0,1}\nonumber \\
  &+\mathcal{T}(k_A \ell)\mathcal{R}(k_B \ell+\varphi)\ket{1,0}
    +\mathcal{T}(k_A \ell)\mathcal{T}(k_B \ell+\varphi)\ket{1,1}.
\end{align}
Its EE is illustrated in Fig. \ref{fig:fig7} as a function of $k_A\ell$,
$k_B\ell$ and $\varphi$.
From the condition for maximal EE, as we found in
Eq. \eqref{eq:maxentcond}, we have
$\left|r_A\right|^2=\left|t_A\right|^2={1}/{2}$, which is obtained for
$k_A\ell=n_{k_A\ell}\pi\pm\arctan\left(2\right)$, together with the
equality $\left|r_{B}r_{B'}^{*}+t_{B}t_{B'}^{*}\right|^{2} = 0$, which
is fulfilled with
$\tan\left(k_{B}\ell\right)\tan\left(k_{B}\ell+\varphi\right)=-4$.
This leads to the solution
\begin{equation}
  \varphi=-\arctan\left(\frac{3\sin\left(2k_{B}\ell\right)}
    {5+3\cos\left(2k_{B}\ell\right)}\right)+\left(2n+1\right)\frac{\pi}{2}.
\end{equation}
When one of the conditions $\left|t_{A}\right|^{2}=0$, or
$\left|r_{A}\right|^{2}=0$, or $\left|r_{B}\right|^{2}=0$,
or $\left|t_{B}\right|^{2}=0$ or $\varphi=2n\pi$ is satisfied, the state
is separable.

\section{Conclusions}

In this letter, we demonstrated a procedure for generating tunable
entanglement using simple quantum graphs via coherent controlled
scattering.
Our results reveal exact analytical conditions for maximal and vanishing
entanglement, depending on the wave number, phase, and topology of each
QG.
Other possible modifications to QGs, such as edge lengths, boundary
conditions on their vertices, or potential along their edges, may serve
to generate entangled states with controllable parameters.
We anticipate that this approach can be implemented using current
microwave network technologies, offering a novel physical platform for
quantum information tasks.

Other investigation possibilities may be related to the connection
with scattering in randomized QGs studied in
Ref. \cite{APQ.1.046126.2024}.
It is also interesting to deal with controlled operations on multiple
edges on the same QG.
Studies involving QGs with three or more leads are also interesting,
as each scattering channel represents an available state of the system.
Thus, a QG with $c$ leads can be used to represent a $c$-level qudit,
with the state now being
$\ket{\Gamma_G^{c}} = S_{\Gamma_G^{c}}\ket{0} =
r_G \ket{0} + \sum_{i=1}^{c-1}t_G^{(i)} \ket{i}$.
Another line of investigation can also include a system of three or more
QGs, as each QG has a Hilbert space, and combining $N$ of them can
be used in the studies of systems with $N$ qubits or even qudits.

In comparison with well-established platforms employed to generate
entanglement, such as optical systems and superconducting qubits, the QG
framework is still less mature experimentally, but it offers
complementary advantages.
In particular, the QG approach differs conceptually from these settings,
as entanglement emerges directly from controlled scattering processes in
simple graph structures, which brings analytical tractability and
flexibility of graph design, and it may find practical realization in
microwave networks.
In this sense, it seems that microwave networks may be a novel source to
for investigating entanglement.
However, to achieve this objective, several practical challenges in
microwave networks must be addressed, including noise mitigation,
precise phase shifts, appropriate wavenumbers, edge lengths, and
boundary conditions, for which the system must be adapted to reliably
perform the intended operation.
We mention that to implement coherent control in microwave networks,
one may treat the devices with a phase shifter
\cite{PRE.107.024203.2023,Nature.574.505.2019} or microwave cavity
\cite{Nature.13.6104.2022} to simulate the required procedure.

\section*{Acknowledgments}
This work was partially supported by Coordenação
de Aperfeiçoamento de Pessoal de Nível Superior (CAPES, Finance Code 001).
It was also supported by Conselho Nacional de Desenvolvimento Científico
Tecnológico (CNPq) and Instituto Nacional de Ciência e Tecnologia de
Informação Quântica (INCT-IQ). DB and FMA acknowledge financial
support from CNPq Grants 303469/2019-6 (DB), 402830/2023-7 (DB), and
313124/2023-0 (FMA), and FAUEPG Project No. 305 (FMA).

\section*{Data Availability Statement}
No Data is associated with the manuscript


\bibliographystyle{iopart-num}
\bibliography{bib.bib}

\end{document}